\documentclass[10pt,superscriptaddress,twocolumn,aps,showpacs,nofootinbib,eqsecnum,twoside,prc]{revtex4}
\usepackage[pdfborder={0 0 0},colorlinks=true,linkcolor=blue,citecolor=blue,filecolor=blue,urlcolor=blue]{hyperref}   
\usepackage{url}       
\usepackage{ulem}
\usepackage{slashed}
\usepackage{amssymb}
\usepackage{amsfonts}
\usepackage{mathbbol} 
\usepackage{amstext}
\usepackage{graphicx}
\usepackage{color}
\usepackage{epsfig,amsmath,lscape}
\usepackage{soul}
\usepackage{mathptmx}                
\usepackage{dcolumn}                 
\usepackage{bm}                      
\usepackage[caption=false]{subfig}
\usepackage{subfig}

\begin{document}

\title{Maxwell equation violation by density dependent magnetic
  fields in neutron stars}

\author{D\'ebora P. Menezes}
\affiliation{Departamento de F\'{\i}sica - CFM - Universidade Federal de Santa Catarina, \\ 
Florian\'opolis - SC - CP. 476 - CEP 88.040 - 900 - Brazil \\ email: debora.p.m@ufsc.br}

\author{Marcelo D. Alloy}
\affiliation{Campus Blumenau- Universidade Federal de Santa Catarina, \\ 
Rua Jo\~ao Pessoa, 2750 - Blumenau - SC - CEP - 89036-002 - Brazil \\ email: marcelo.alloy@ufsc.br}

\begin{abstract}
We show that the widely used density dependent magnetic field prescriptions,
necessary to account for the variation of the field intensity from the
crust to the core of neutron stars violate one of the Maxwell
equations. We estimate how strong the violation is when different
equations of state are used and check for
which cases the pathological problem can be cured.
\end{abstract}

\pacs {26.60.-c, 97.60.Jd, 11.10.-z}

\maketitle


\section{Motivation}

\paragraph*{}
The physics underlying the quantum chromodynamics (QCD) phase diagram has
still not been probed at all temperatures and densities. While some
aspects can be confirmed either by lattice QCD or experimentally in
heavy ion collisions for instance, other aspects depend on
extraterrestrial information. One of them is the possible constitution
and nature of neutron stars (NS), which are compact objects related to the low
temperature and very high density portion of the QCD phase diagram.
Astronomers and astrophysicists can provide a handful of information
obtained from observations and infer some macroscopic properties, namely NS
masses, radii, rotation period and external magnetic fields, which have been guiding
the theory involving microscopic equations of state (EOS) aimed to describe
this specific region of the QCD phase diagram.

\paragraph*{}
There are different classes of NS and three of them have shown to be
compatible with highly magnetised compact objects, known as magnetars
\cite{Duncan, Usov}, namely, the soft gamma-ray repeaters, the anomalous 
X-ray pulsars and more recently, the repeating fast radio burst \cite{frb}.
The quest towards explaining these NS with strong surface magnetic fields,
 has led to a prescription that certainly violates Maxwell 
equations \cite{Chakrabarty}. The aim of this letter is to
show how strong this violation is and check whether the density dependent
magnetic field  prescriptions can or cannot be justified.
Magnetars are likely to bear magnetic fields of the order of
$10^{15}$ G on their surfaces, which are three orders of magnitude
larger than magnetic fields in standard NS. In the last years,
many papers dedicated to the study of these objects have shown
that the EOS are only sensitive to magnetic fields as large as $10^{18}$ G
or stronger \cite{originalB,originalNS}. The Virial theorem and the fact that
some NS can be quark (also known as strange) stars allow these objects to
support central magnetic fields as high as $3 \times 10^{18}$ G  if they
contain  hadronic constituents and up to $10^{19}$ G if they are quark
stars. To take into account these
possibly varying magnetic field strength that increases towards the
centre of the stars, the following proposition was made \cite{Chakrabarty}:
\begin{equation}
 B_z(n) = B^{surf} + B_0\bigg [ 1 - \exp \bigg \{ - \beta \bigg (
 \frac{n}{n_0} \bigg )^{\gamma} \bigg \} \bigg ], 
\label{brho} 
\end{equation}
where  $B^{surf}$  is the magnetic field on the surface of the neutron
stars taken as $10^{14}G$ in the original paper, $n$ is the total number density, and $n_0$ is the nuclear saturation density
In subsequent papers \cite{Mao,Rabhi,Menezes1,Ryu,Rabhi2,Mallick,Lopes1,Dex,Benito1,Benito2,Mallick2,Ro,Dex2}, the above prescription was extensively used,
with many variations in the values of  $B^{surf}$, generally taken as
$10^{15}$G on the surface, $\beta$ and
$\gamma$, arbitrary parameters that cannot be tested by astronomical
observations. The high degree of arbitrariness was checked already in
\cite{Chakrabarty} and later in \cite{Lopes2015} and more than 50\%
variation in the maximum stellar mass and 25\% variation in the
corresponding radius was found.

\paragraph*{}
With the purpose of reducing the number of free parameters from two to
one and consequently the arbitrariness in the results, another prescription was
then proposed in \cite{Lopes2015}:
\begin{equation}
B_z(\epsilon) = B^{surf} +  B_0 \bigg ({\frac{\epsilon}{\epsilon_c}} \bigg
)^{\alpha},
\label{bepsilon}
\end{equation}
where  $\epsilon_c$  is the energy density at the centrer of the maximum mass neutron star
with zero magnetic field, $\alpha$ is any positive number and
$B_0$ is the fixed value of the magnetic field. With this recipe, all 
magnetic fields converge to a certain value at some large energy
density, despite the $\alpha$ value used. 

\paragraph*{}
One of the Maxwell equations tells us that $\nabla\cdot\vec B=0$. In most 
neutron star calculations, the magnetic field is chosen as static and
constant in the $z$ direction, as proposed in the first application to
quark matter \cite{originalB}. In this case, the energy associated
with the circular motion in the $x-y$ plane is quantised (in units of
$2qB$, $q$ being the electric charge) and the energy along $z$ is
continuous. The desired EOS is then obtained and the energy density,
pressure and number density depend on the filling of the Landau
levels. Apart from the original works \cite{originalB, originalNS},
detailed calculations for different models can also be found, for
instance, in \cite{Menezes1,Benito1} and we will not enter into
details here. However, it is important to stress that the magnetic
field is taken as constant in the $z$ direction, what results 
in different contributions in the $r$ and $\theta$ directions when one
calculates the magnetic field in spherical coordinates. 

\paragraph*{}
According to \cite{Goldreich,Danielle}, and assuming a perfectly
conducting neutron star ($B_r(R)=0$) that bears a magnetic dipole moment aligned
with the rotation axis such that $\mu=B_p R^3/2$, where $R$ is the
radius of the star and $B_p$ the magnetic field intensity at the pole,
the components of the magnetic field in spherical coordinates are given
by:
\begin{equation}
B_r=B_P \cos \theta \left(\frac{R}{r}\right)^3, \quad
B_\theta=\frac{B_P}{2}  \sin \theta \left(\frac{R}{r}\right)^3,
\label{daniele}
\end{equation}
and in this case, it is straightforward to show that $\nabla \cdot \vec
B=0$. 

\paragraph*{}
If one cast eqs. (\ref{brho}) and (\ref{bepsilon}) in spherical
coordinates, from now on called respectively original and LL's prescriptions,
they acquire the form
$B_r=\cos \theta B_z$, $B_\theta=-\sin \theta B_z$ and $B_\phi=0$
and the resulting divergent reads 
\begin{equation}
\nabla \cdot \vec B= \cos \theta \frac{\partial B}{\partial r},
\label{divB}
\end{equation}
where the magnetic field in the radial direction can be obtained from the
solution of the TOV equations \cite{tov}, where $r$ runs from the centre
to the radius of the star. As a simple conclusion, $\nabla\cdot\vec B$ is
generally not zero, except for some specific values of the parameters
that we discuss in the next Section.

\section{Results and Discussion}

\paragraph*{}
In what follows we analyse how much $\nabla\cdot\vec B$ deviates from zero when $B$
is allowed to vary either with the original prescription as in
eq. (\ref{brho}) or with LL's proposal, as in eq. (\ref{bepsilon})
with two different models, the NJL and the MIT bag model.
These two models have been extensively used to describe stellar matter
in the interior of quark stars. It is important to point out that the
same test could be performed with hadronic models used to account for
magnetised NS with hadronic constituents, as in \cite{
Mao,Rabhi,Ryu,Rabhi2,Mallick,Lopes1,Dex,Benito1,Mallick2,Ro}, for
instance.

\begin{figure}[ht]
\includegraphics[scale=0.40]{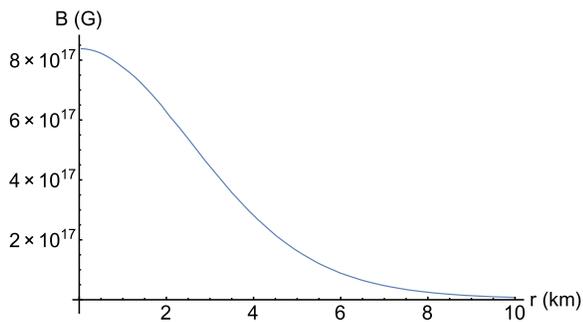}
\caption{ Magnetic field versus r  for the MIT model, bag constant
148 MeV$^{1/4}$, $M_{max}=1.4~M_\odot$ and $R=10.04$ km.}
\label{fig0}
\end{figure}

\paragraph*{}
Before we analyse the behaviour of $\nabla\cdot\vec B /B$ that depends
on $\frac{\partial B}{\partial r} = \frac{\partial B}{\partial n}
\frac{\partial n}{\partial r}$,  we show in
Fig. \ref{fig0} how the magnetic field varies with the star radius for
one specific case, i.e., the MIT bag model with $B=10^{18}$ G and the prescription given
in eq.(\ref{bepsilon}). All other cases studied next present a very similar
behaviour. It is interesting to notice that the curve changes
concavity around half the stellar radius. 

\begin{figure}[ht]
\includegraphics[scale=0.44]{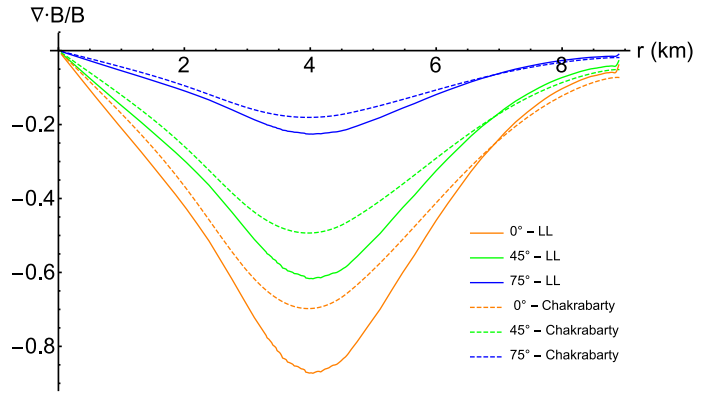}
\caption{Equations of state obtained with the NJL model and
  $B_0=10^{18}$ G. Chakrabarty's prescription was calculated with 
$M_{max}=1.44~M_\odot$ and $R=8.88$ km, $\beta=5 \times 10^{-4}$ and $\gamma=3$.
LL's prescription was calculated with 
$M_{max}=1.46~M_\odot$, $R=8.83$ km and $\zeta=3$.}
\label{fig1}
\end{figure}

\paragraph*{}
In Fig. \ref{fig1} we plot $\nabla\cdot\vec B /B$ as a function of the star
radius for different latitude ($\theta$) angles and a magnetic field
equal to $B_0=10^{18} G$. In both cases
the equations of state were obtained with the Nambu-Jona-Lasinio
model \cite{Menezes1, Luiz2016}. The violation is quite strong and
$\nabla\cdot\vec B /B$ can reach 70\% for small angles.

\begin{figure}[ht]
\includegraphics[scale=0.39]{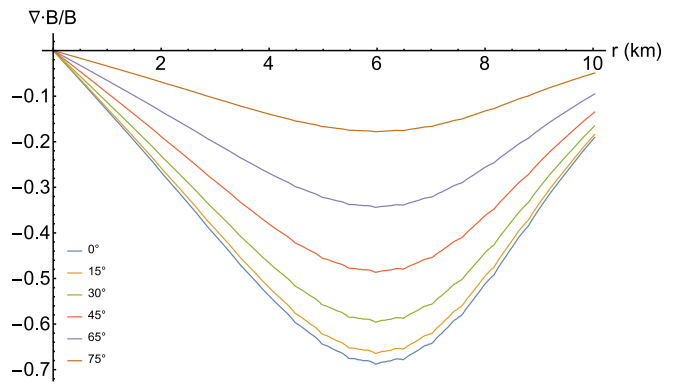}
\caption{Equations of state obtained with the MIT model, bag constant
148 MeV$^{1/4}$, $M_{max}=1.4~M_\odot$ and $R=10.04$ km for
  different latitude angles. }
\label{fig2}
\end{figure}

\begin{figure}[ht]
\includegraphics[scale=0.5]{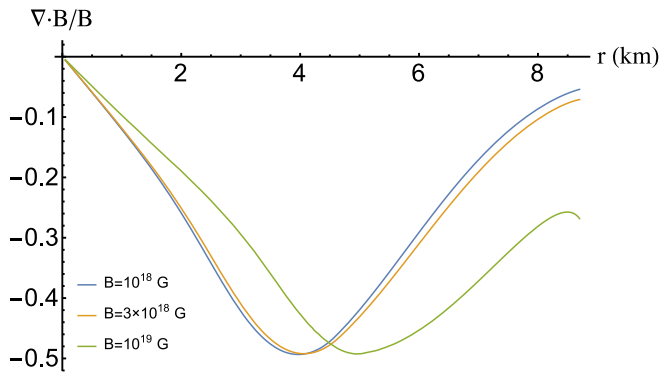}
\caption{ Quark stars described by NJL model with different values of
  $B_0$ and $\theta=45$ degrees.
For $B_0=10^{18}$ G, $M_{max}=1.44_\odot$, $R=8.88$ km and central
energy density $\epsilon_c=7.67~fm^{-4}$. For $B_0=3\times10^{18}$ G:
$M_{max}=1.45_\odot$, $R=8.88$ km and central energy density
$\epsilon_c=7.65~fm^{-4}$. For $B_0=10^{19} G$: $M_{max}=1.50_\odot$,
$R=8.80$ km and central energy density $\epsilon_c=8.11~fm^{-4}$. }
\label{fig3}
\end{figure}

\paragraph*{}
In Figure \ref{fig2} we plot the same quantity as in Figure \ref{fig1}
for the LL's prescription and the much simpler and also more used MIT
bag model for different latitude angles. Again the violation amounts
to the same values as the ones obtained within the NJL model.
 Finally, in Figure \ref{fig3}, we show how large  the deviation can be
for different values of the magnetic field intensity and a fixed angle
$\theta=45$ degrees and the original prescription. We see that the
deviation reaches approximately the same percentages, independently of
the field intensity.

\paragraph*{}
Now, we turn our attention to a possible
generalisation of eqs. (\ref{brho}) and (\ref{bepsilon}) in spherical coordinates
in order to verify for which conditions the divergent becomes zero
and check whether the situation can be circumvented. The
generalized magnetic field components is given by
\begin{equation}
B_r= B_0 \cos \theta \left( f(r)\right)^\eta, \quad
B_\theta=-\frac{B_0}{\zeta} \sin \theta \left( f(r)\right)^\eta,
\end{equation}
and the corresponding divergent reads:
\begin{equation}
\nabla \cdot \vec B= B_0 \cos \theta \left[ 
\frac{2 f(r)^{\eta}}{r} + \eta f(r)^{\eta-1} \frac{df}{dr} -
\frac{2 f(r)^{\eta}}{r \zeta} \right], 
\label{divBLL}
\end{equation}
which is zero either for the trivial solution $\cos \theta=0$ or if
$$\frac{2}{r} + \frac{\eta}{f(r)} \frac{df}{dr} -
\frac{2}{r \zeta}=0, $$ 
for which a general solution has the form
$f(r)=A r^{\frac{2-2 \zeta}{\eta \zeta}}$ . When $\zeta=-2$ and
  $\eta=3$, $A=R$ and $B_0=B_p$, eqs.(\ref{daniele}) are recovered.
When $\zeta=1$, the numerator of the exponent becomes zero and $f(r)$
is simply a constant ($A$), does not depending on $r$. 
Another possibility is the assumption that $f(r)$ is a function of the
density, for instance, as $f(n(r))$ or of the energy density as 
$f(\varepsilon(r))$. 
In these cases, 
$$ \frac{df}{d n} \frac{d n}{dr} = \frac{2~f}{r~\eta} \left(
  \frac{1-\zeta}{\zeta}\right), \quad
\frac{df}{d \varepsilon} \frac{d \varepsilon}{dr} = \frac{2~f}{r~\eta} \left(
  \frac{1-\zeta}{\zeta}\right).$$

If we take $\zeta=1$, the result resembles the original (LL's) prescription, 
$ \frac{df}{d n} =0$ ($\frac{df}{d \varepsilon} =0$)
 because $\frac{d n}{dr}$ ($\frac{d \varepsilon}{dr}$) is obtained from
the TOV equations and is never zero. For $\zeta \ne 1$, solutions can
be obtained from numerical integration.
\\
\\

\section{Final Remarks}

\paragraph*{}
We have shown that both existent prescriptions for a density dependent 
magnetic field widely used in the study of magnetised neutron star
matter equations of state strongly violate one of the Maxwell
equations, showing a pathological problem.
However, they can be cured if used only for
well defined values of functionals that  guarantee
that the Maxwell equation is not violated.
We have shown the results obtained from EOS used to describe quarks
stars. Had we shown the same quantities for neutron stars constituted
of hadronic matter only, the qualitative results would be the same.
As a final word of caution, we would like to comment that the
divergence problem in general relativity goes beyond the simple
analysis we have performed and will deserve an attentive look in the future.

\begin{acknowledgments}
 This work was partially supported by CNPq under grant 300602/2009-0.
We would like to thank very fruitful discussions with Prof. Constan\c
ca Provid\^encia.
\end{acknowledgments}


\begin{thebibliography}{99}

\bibitem{Duncan} R.~C.~Duncan, C.~Thompson, 
Astrophys. J. $\textbf{392}$, L9 (1992);
R.~C.~Duncan, C.~Thompson, 
Mon. Not. R. Astron. Soc. $\textbf{275}$, 255 (1995); R.~C.~Duncan,
C.~Thompson, 
Astrophys. J. $\textbf{473}$, 322 (1996).

\bibitem{Usov} V.~V.~Usov, 
Nature $\textbf{357}$, 472 (1992).

\bibitem{frb} L.G. Spitler et al, Nature 531, 202 (2016).

\bibitem{Chakrabarty} D.~Bandyopadhyay, S.~Chakrabarty, S.~Pal,
Phys. Rev. Lett. $\textbf{79}$, 2176 (1997)

\bibitem{originalB} S. Chakrabarty, Phys. Rev. D 54, 1306 (1996).

\bibitem{originalNS} A. Broderick, M. Prakash, and J. M. Lattimer, Astrophys. J. 537,
351 (2000).

\bibitem{Mao}G.~J.~Mao, C.~J.~Mao, A.~Iwamoto, Z.~X.~Li, 
Chin. J. Astron. Astrophys. $\textbf{3}$, 359 (2003).

\bibitem{Rabhi} A.~Rabhi et al.,
J. Phys. G $\textbf{36}$, 115204 (2009). 

\bibitem{Menezes1} D.~P.~Menezes et al, Phys. Rev. C, 79, 035807 (2009);
D.~P.~Menezes et al., 
Phys. Rev. C $\textbf{80}$, 065805 (2009).

\bibitem{Ryu} C.~Y.~Ryu, K.~S.~Kim, M.~Ki Cheoun, 
Phys. Rev. C $\textbf{82}$, 025804 (2010).

\bibitem{Rabhi2}A.~Rabhi, P.~K.~Panda, C.~Providencia, 
Phys. Rev. C $\textbf{84}$, 035803 (2011).

\bibitem{Mallick}R.~Mallick, M.~Sinha, 
Mon. Not. R. Astron. Soc. $\textbf{414}$, 2702 (2011).

\bibitem{Lopes1}L.~L.~Lopes, D.~P.~Menezes, 
Braz. J. Phys. $\textbf{42}$, 428 (2012).

\bibitem{Dex}V.~Dexheimer, R.~Negreiros, S.~Schramm, 
Eur. J. Phys. A $\textbf{48}$, 189 (2012) 

\bibitem{Benito1}R.~H.~Casali, L.~B.~Castro, D.~P.~Menezes, 
Phys. Rev. C $\textbf{89}$, 015805 (2014).

\bibitem{Benito2}D.~P.~Menezes et al., 
Phys. Rev. C $\textbf{89}$ 055207 (2014).

\bibitem{Mallick2} R.~Mallick, S.~Schramm, 
Phys. Rev. C $\textbf{89}$, 045805 (2014).

\bibitem{Ro}R.~O.~Gomes, V.~Dexheimer, C.~A.~Z.~Vasconcellos, 
Astron. Nachr. $\textbf{335}$, 666 (2014).

\bibitem{Dex2}V.~Dexheimer, D.~P.~Menezes M.~Strickland, 
J. Phys. G $\textbf{41}$, 015203 (2014).

\bibitem{Lopes2015} Luiz Lopes and Debora Menezes,  J. Cosmology and
Astroparticle Physics 08 (2015) 002.

\bibitem{Danielle} Daniele Vigan\`o, {\it Magnetic Fields in Neutron
    Stars}, Ph.D. Thesis, Universidad de Alicante, 2013.

\bibitem{Goldreich} P. Goldreich and W.H. Julian, ApJ 157, 869 (1969).

\bibitem{tov}  R.C. Tolman, Phys. Rev. {\bf 55}, 364 (1939);
J.~R.~Oppenheimer, G.~M.~Volkoff, 
Phys. Rev. $\textbf{55}$, 374 (1939)

\bibitem{Luiz2016} D\'ebora Peres Menezes and Luiz La\'ercio Lopes, 
Eur. Phys. J. A (2016) 52:17.

\end{thebibliography}
\end{document}